\documentclass{article}
\def\~{\tilde}
\usepackage{graphicx}

\begin{document}
\title{Bohmian prediction about a two double-slit experiment and its disagreement with standard quantum mechanics}
\author{M. Golshani \footnote{E-mail: golshani@ihcs.ac.ir} and O. Akhavan \footnote{E-mail:
akhavan@mehr.sharif.edu }}

\date{\small{{\it Department of Physics, Sharif University of
Technology, P.O. Box 11365--9161, Tehran,  Iran\\Institute for
Studies in Theoretical Physics and Mathematics, P.O. Box
19395--5531, Tehran, Iran}}} \maketitle

\begin{abstract}
The significance of proposals that can predict different results
for standard and Bohmian quantum mechanics have been the subject
of many discussions over the years.  Here, we suggest a particular
experiment (a two double-slit experiment) and a special detection
process, that we call selective detection, to distinguish between
the two theories.  Using our suggested experiment, it is shown
that the two theories predict different observable results at the
individual level for a geometrically symmetric arrangement.
However, their predictions are the same at the ensemble level.  On
the other hand, we have shown that at the statistical level, if we
use our selective detection, then either the predictions of the
two theories differ or where standard quantum mechanics is silent
or vague, Bohmian quantum mechanics makes explicit predictions.
\end{abstract}

PACS number(s): 03.65.Bz\\

\section{Introduction}
According to the standard quantum mechanics (SQM), the complete
description of a system of particles is provided by its
wavefunction.  The empirical predictions of SQM follow from a
mathematical formalism which makes no use of the assumption that
matter consists of particles pursuing  definite tracks in
space-time.  It follows that the results of the experiments
designed to test the predictions of the theory, do not permit us
to infer any statement regarding the particle--not even its
independent existence.  But, in the Bohmian quantum mechanics
(BQM), the additional element that is introduced apart from the
wavefunction is the particle position, conceived in the classical
sense as pursuing a definite continuous track in space-time [1-3].
The detailed predictions made by this causal interpretation
explains how the results of quantum experiments come about but it
is claimed that they are not tested by them. In fact when Bohm
{\cite{Bohm}} presented his theory in 1952, experiments could be
done with an almost continuous beam of particles, but not with
individual particles. Thus, Bohm constructed his theory in such a
fashion that it would be impossible to distinguish observable
predictions of his theory from SQM.  This can be seen from Bell's
comment about empirical equivalence of the two theories when he
said:``{\it{It}} [the de Broglie-Bohm version of non-relativistic
quantum mechanics] {\it{is experimentally equivalent to the usual
version insofar as the latter is unambiguous}}"{\cite{Bell}}.
However, could it be that a certain class of phenomena might
correspond to a well-posed problem in one theory but to none in
the other? Or might the additional particles and definite
trajectories of Bohm's theory lead to a prediction of an
observable where SQM would just have no definite prediction to
make? To draw discrepancy from experiments involving the particle
track, we have to argue in such a way that the observable
predictions of the modified theory are in some way functions of
the trajectory assumption.  The question raised here is whether
the de Broglie-Bohm particle law of motion can be made relevant to
experiment.  At first, it seems that definition of time spent by a
particle within a classically forbidden barrier provides a good
evidence for the preference of BQM.  But, there are difficult
technical questions, both theoretically and experimentally, that
are still unsolved about this tunnelling times {\cite{Cushing}}. A
recent work indicates that it is not practically feasible to use
tunnelling effect to distinguish between the two theories
{\cite{Abolhasani}}.

 On the other hand, Englert {\it{et al.}} {\cite{Englert}} and Scully {\cite{Scully}} have claimed that in some cases Bohm's approach
gives results that disagree with those obtained from SQM and, in
consequence, with experiment.  Again, at first Dewdney {\it{et
al.}} {\cite{Dewdney}} and then Hiley {\it{et al.}} {\cite{Hiley}}
showed that the specific objections raised by Englert {\it{et
al.}} and Scully cannot be sustained. Furthermore, Hiley believes
that no experiment can decide between the standard interpretation
and Bohm's interpretation. However, Vigier {\cite{Vigier}}, in his
recent work, has given a brief list of new experiments which
suggest that the U(1) invariant massless photon assumed properties
of light within the standard interpretation, are too restrictive
and that the O(3) invariant massive photon causal de Broglie-Bohm
interpretation of quantum mechanics, is now supported by
experiments.  Furthermore, in some of the recent investigation,
some feasible experiments have been suggested to distinguish
between SQM and BQM {\cite{Ghose,Golshani}}. In one work, Ghose
indicated that although BQM is equivalent to SQM when averages of
dynamical variables are taken over a Gibbs ensemble of Bohmian
trajectories, the equivalence breaks down for ensembles built over
clearly separated short intervals of time in specially entangled
two-bosonic particle systems {\cite{Ghose}}.  Another one
{\cite{Golshani}} is an extension of Ghose's work to show
disagreement between SQM and BQM in a two-particle system with an
unentangled wavefunction, particularly at the statistical
level\footnote{To clarify our discussion it is worth noting that
in this paper we have used the following definitions:\\** By
statistically level we mean our final interference pattern.\\***
The individual level refers to our experiment with a pair of
particles which are emitted in clearly separated short intervals
of time.}.  Further discussion of this subject can be found in
[13-15]. In that experiment, to obtain a different interference
pattern from SQM, we must deviate the source from its
geometrically symmetric location.

In this investigation, we are offering a new thought experiment
which can decide between SQM and BQM.  Here, the deviation of the
source from its geometrical symmetric location is not necessary
and we have used a system consisting two correlated particles with
an entangled wavefunction.

In the following section, we have introduced a two double-slit
experimental set-up.  In section 3, Bohm's interpretation is used
to find some observable results about our suggested experiment.
Predictions of the standard interpretation and their comparison
with Bohmian predictions is examined in section 4.  In section 5,
we have used selective detection and have compared SQM and BQM
with our thought experiment at the ensemble level of particles,
and we state our conclusion in section 6.

\begin{figure}[t]
\includegraphics[width=15cm,height=8cm,angle=0]{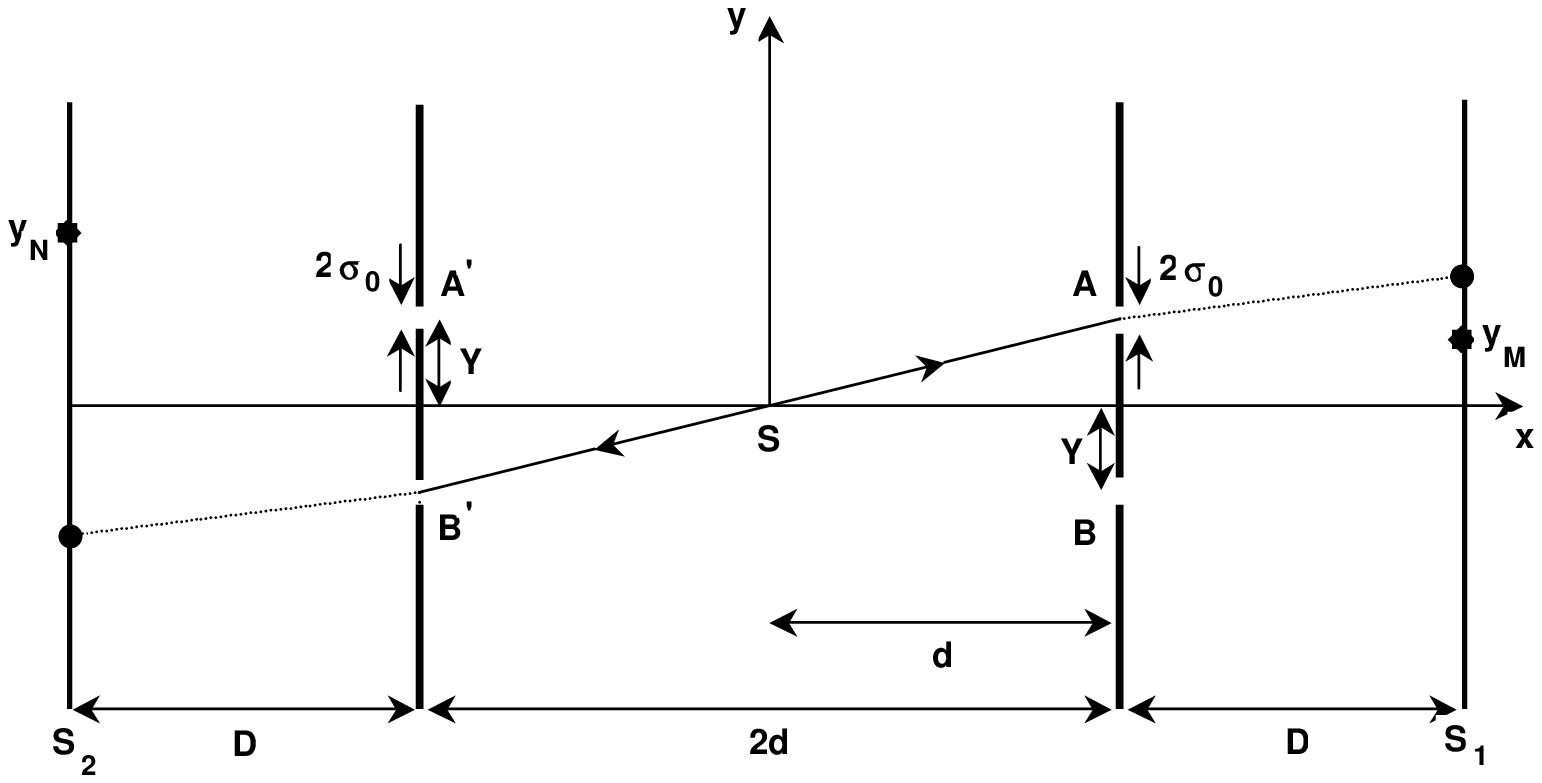}
\caption{A two double-slit experiment configuration. Two identical
particles with zero total momentum are emitted from the source $S$
and then they pass through slits $A$ and $B^{'}$ or $B$ and
$A^{'}$.  Finally, they are detected on $S_{1}$ and $S_{2}$
screens, simultaneously.  It is necessary to note that dotted
lines are not real trajectories.}
\end{figure}
\section{Two double-slit experiment presentation}
To distinguish between SQM and BQM we consider the following
experimental set-up.  A pair of identical non-relativistic
particles with total momentum zero labelled by 1 and 2, originate
from a point source S that is placed exactly in the middle of a
two double-slit screens as shown in Fig. 1.  We assume that the
intensity of the beam is so low that during any individual
experiment we have only a single pair of particles passing through
the slits and the detectors have the opportunity to relate
together for performing selective detection process.  In addition,
we assume that the detection screens $S_{1}$ and $S_{2}$ register
only those pairs of particles that reach the two screens
simultaneously.  Thus, we are sure that the registration of single
particles is eliminated from final interference pattern. The
detection process at the screens $S_{1}$ and $S_{2}$ may be
nontrivial but they play no causal role in the basic phenomenon of
the interference of particles waves {\cite{Holland}}.  In the
two-dimensional system of coordinates $(x,y)$ whose origin $S$ is
shown, the center of slits lie at the points $(\pm d,\pm Y)$. The
wave incident on the slits will be taken as a plane of the form
\begin{equation}
\psi_{in}(x_{1},y_{1};x_{2},y_{2};t)=ae^{i[k_{x}(x_{1}-x_{2})+k_{y}(y_{1}-y_{2})]}e^{-iEt/\hbar}
\end{equation}
where $a$ is a constant and
$E=E_{1}+E_{2}=\hbar^{2}(k_{x}^{2}+k_{y}^{2})/m$ is the total
energy of the system of the two particles.  The plane wave
assumption comes from large distance between source $S$ and
double-slit screens.  To avoid the mathematical complexity of
Fresnel diffraction at a sharp-edge slit, we suppose the slits
have soft edges that generate waves having identical Gaussian
profiles in the $y$-direction while the plane wave in the
$x$-direction is unaffected {\cite{Holland}}.  The instant at
which the packets are formed will be taken as our zero of time.
Therefore, the four waves emerging from the slits $A$, $B$,
$A^{'}$ and $B^{'}$ are initially
\begin{equation}
\psi_{A,B}(x,y)=a(2\pi\sigma_{0}^{2})^{-1/4}e^{-(\pm
y-Y)^{2}/4\sigma_{0}^{2}}e^{i[k_{x}(x-d)+k_{y}(\pm y-Y)]}
\end{equation}
\begin{equation}
\psi_{A^{'},B^{'}}(x,y)=a(2\pi\sigma_{0}^{2})^{-1/4}e^{-(\pm
y+Y)^{2}/4\sigma_{0}^{2}}e^{i[-k_{x}(x+d)+k_{y}(\pm y+Y)]}
\end{equation}
where $\sigma_{0}$ is the half-width of each slit.  At time $t$
the general total wavefunction at a space point $(x,y)$ of our
considered system for bosonic and fermionic particles is given by
\begin{eqnarray}
\psi(x_{1},y_{1};x_{2},y_{2};t)=N[\psi_{A}(x_{1},y_{1},t)\psi_{B^{'}}(x_{2},y_{2},t)\pm\psi_{A}(x_{2},y_{2},t)\psi_{B^{'}}(x_{1},y_{1},t)\nonumber\\
 +\psi_{B} (x_{1},y_{1},t)\psi_{A^{'}}(x_{2},y_{2},t)\pm\psi_{B} (x_{2},y_{2},t)\psi_{A^{'}}
(x_{1},y_{1},t)]
\end{eqnarray}
with
\begin{equation}
\psi_{A,B}(x,y,t)=a(2\pi \sigma_{t}^{2})^{-1/4}e^{-(\pm
y-Y-u_{y}t)^{2}/4\sigma_{0}\sigma_{t}}e^{i[k_{x}(x-d)+k_{y}(\pm
y-Y-u_{y}t/2)-E_{x}t/\hbar]}
\end{equation}
\begin{equation}
\psi_{A^{'},B^{'}}(x,y,t)=a(2\pi \sigma_{t}^{2})^{-1/4}e^{-(\pm
y-Y-u_{y}t)^{2}/4\sigma_{0}\sigma_{t}}e^{i[-k_{x}(x+d)+k_{y}(\pm
y-Y-u_{y}t/2)-E_{x}t/\hbar]}
\end{equation}
where $N$ is a reparametrization constant that its value is
unimportant in this paper and
\begin{eqnarray}
\sigma_{t}=\sigma_{0}(1+\frac{i\hbar t}{2m\sigma_{0}^{2}})\\
u_{y}=\frac{\hbar k_{y}}{m}\nonumber\\
E_{x}=\frac{1}{2}mu_{x}^{2}
\end{eqnarray}
where $u_{x}$ and $u_{y}$, according to BQM, are initial group
velocities corresponding to each particle in the $x$- and
$y$-directions, respectively.  In addition, the upper and lower
sings in the total wavefunction refer to symmetric and
anti-symmetric wavefunction under exchange of particle 1 to
particle 2, corresponding to bosonic and fermionic property, while
in the equations (5) and (6) they refer to upper and lower slits,
respectively.  In the next section, we have used BQM to drive some
results of this experiment.

\section{Bohmian predictions about the suggested experiment}

In BQM, the complete description of a system is given by
specifying the position of the particles in addition to their
wavefunction which has the role of guiding the particles according
to following guidance condition for $n$ particles, with masses
$m_1, m_2,..., m_n$
\begin{equation}
\overrightarrow{\dot{x}_{i}}(\overrightarrow{x},t)=\frac{1}{m_{i}}
\overrightarrow{\nabla_{i}}S(\overrightarrow{x},t)=\frac{\hbar}{m_{i}}Im\left(
\frac{\overrightarrow{\nabla_{i}}\psi(\overrightarrow{x},t)}{\psi(\overrightarrow{x},t)}\right)
\end{equation}
where $\overrightarrow{x}=(\overrightarrow{x_1}
,\overrightarrow{x_2},..., \overrightarrow{x_n})$ and
\begin{equation}
\psi(\overrightarrow{x_1},\overrightarrow{x_2},...,\overrightarrow{x_n};t)=
R(\overrightarrow{x_1},\overrightarrow{x_2},...,\overrightarrow{x_n};t)
e^{iS(\overrightarrow{x_1},\overrightarrow{x_2},...,\overrightarrow{x_n};t)/\hbar}
\end{equation}
is a solution of Schr$\ddot{o}$dinger's wave equation.  Thus,
instead of SQM with indistinguishable particles, in BQM the path
of particles or their individual histories distinguishes them and
each one of them can be studied separately {\cite{Holland}}.  In
addition, Belousek {\cite{Belousek}} in his recent work, concluded
that the problem of Bohmian mechanical particles being
statistically (in)distinguishable is a matter of theory choice
underdetermined by logic and experiment, and that such particles
are in any case physically distinguishable.  For our considered
experiment, the speed of the particles 1 and 2 in the
$y$-direction is given , respectively, by
\begin{equation}
\dot{y}_{1}(x_{1},y_{1};x_{2},y_{2};t)=\frac{\hbar}{m}Im\frac{\partial_{y_{1}}\psi(x_{1},y_{1};x_{2},y_{2};t)
}{\psi(x_{1},y_{1};x_{2},y_{2};t)}
\end{equation}
\begin{equation}
\dot{y}_{2}(x_{1},y_{1};x_{2},y_{2};t)=\frac{\hbar}{m}Im\frac{\partial_{y_{2}}\psi(x_{1},y_{1};x_{2},y_{2};t)
}{\psi(x_{1},y_{1};x_{2},y_{2};t)}.
\end{equation}
With the replacement of $\psi(x_{1},y_{1};x_{2},y_{2};t)$ from
(4), (5) and (6), we have
\begin{eqnarray}
\dot{y}_{1}&=&N\frac{\hbar}{m}Im\{\frac{1}{\psi}[[-2(y_{1}-Y-u_{y}t)/4\sigma_{0}\sigma_{t}+ik_{y}]\psi_{A_{1}}\psi_{B^{'}_{2}}\cr
           &\pm&[-2(y_{1}+Y+u_{y}t)/4\sigma_{0}\sigma_{t}-ik_{y}]\psi_{A_{2}}\psi_{B^{'}_{1}}\cr
           &+&[-2(y_{1}+Y+u_{y}t)/4\sigma_{0}\sigma_{t}-ik_{y}]\psi_{B_{1}}\psi_{A^{'}_{2}}\cr
           &\pm&[-2(y_{1}-Y-u_{y}t)/4\sigma_{0}\sigma_{t}+ik_{y}]\psi_{B_{2}}\psi_{A^{'}_{1}}]\}
\end{eqnarray}
\begin{eqnarray}
\dot{y}_{2}&=&N\frac{\hbar}{m}Im\{\frac{1}{\psi}[[-2(y_{2}+Y+u_{y}t)/4\sigma_{0}\sigma_{t}-ik_{y}]\psi_{A_{1}}\psi_{B^{'}_{2}}\cr
           &\pm&[-2(y_{2}-Y-u_{y}t)/4\sigma_{0}\sigma_{t}+ik_{y}]\psi_{A_{2}}\psi_{B^{'}_{1}}\cr
           &+&[-2(y_{2}-Y-u_{y}t)/4\sigma_{0}\sigma_{t}+ik_{y}]\psi_{B_{1}}\psi_{A^{'}_{2}}\cr
           &\pm&[-2(y_{2}+Y+u_{y}t)/4\sigma_{0}\sigma_{t}-ik_{y}]\psi_{B_{2}}\psi_{A^{'}_{1}}]\}
\end{eqnarray}
where, for example, the short notation
$\psi_{A}(x_{1},y_{1},t)=\psi_{A_{1}}$ is used. Furthermore, from
(5) and (6) it is clear that
\begin{eqnarray}
&&\psi_{A}(x_{1},y_{1},t)=\psi_{B}(x_{1},-y_{1},t)\cr
&&\psi_{A}(x_{2},y_{2},t)=\psi_{B}(x_{2},-y_{2},t)\cr
&&\psi_{B^{'}}(x_{1},y_{1},t)=\psi_{A^{'}}(x_{1},-y_{1},t)\cr
&&\psi_{B^{'}}(x_{2},y_{2},t)=\psi_{A^{'}}(x_{2},-y_{2},t)
\end{eqnarray}
 which indicates the reflection symmetry  of $\psi(x_{1},y_{1};x_{2},y_{2};t)$ with
respect to the $x$--axis.  Utilizing this symmetry in (13) and
(14), we can see that
\begin{eqnarray}
&&\dot{y}_{1}(x_{1},y_{1};x_{2},y_{2};t)=-\dot{y}_{1}(x_{1},-y_{1};x_{2},-y_{2};t)\cr&&
\dot{y}_{2}(x_{1},y_{1};x_{2},y_{2};t)=-\dot{y}_{2}(x_{1},-y_{1};x_{2},-y_{2};t)
\end{eqnarray}
which are valid  for both bosonic and fermionic particles.
Relations (16) show that if $y_{1}(t)=y_{2}(t)=0$, then the speed
of each particles in the $y$-direction is zero along the symmetry
axis $x$. This means that none of the particles can cross the
$x$-axis nor are they tangent to it, provided both of them are
simultaneously on this axis.  Similar conclusions can be found in
some other works [for example, 8, 9, 11-13].  It can be seen that
there is the same symmetry of the velocity about the $x$-axis as
for an ordinary double-slit experiment {\cite{Holland}}.

If we consider $y=(y_{1}+y_{2})/2$ to be the vertical coordinate
of the center of mass of the two particles, then we can write
\begin{eqnarray}
\dot{y}&=&(\dot{y}_{1}+\dot{y}_{2})/2\cr
       &=&-N\frac{\hbar}{2m}Im\{{\frac{1}{\psi}(\frac{y_{1}+y_{2}}{2\sigma_{0}\sigma_{t}})
       (\psi_{A_{1}}\psi_{B^{'}_{1}}\pm\psi_{A_{2}}\psi_{B^{'}_{1}}+\psi_{B_{1}}\psi_{A^{'}_{2}}\pm\psi_{B_{2}}\psi_{A^{'}_{1}})}\}\cr
       &=&\frac{(\hbar/2m\sigma_{0}^{2})^{2}}{1+(\hbar/2m\sigma_{0}^{2})^{2}t^{2}}yt
       .
\end{eqnarray}
Solving the equation of motion (17), we obtain the path of the
$y$-coordinate of the center of mass
\begin{equation}
y=y_{0}\sqrt{1+(\hbar/2m\sigma_{0}^{2})^{2}t^{2}}.
\end{equation}
If at $t=0$ the center of mass of the two particles is exactly on
the $x$-axis, then $y_{0}=0$, and the center of mass of the
particles will always remain on the $x$-axis. Thus, according to
BQM, the two particles will be detected at points symmetric with
respect to the $x$-axis, as shown in Fig. 1.

It seems that calculation of quantum potential can give us another
perspective of this experiment.  As we know, to see the connection
between the wave and particle, the Schr$\ddot{o}$dinger equation
can be rewritten in the form of a generalized Hamilton-jacobi
equation that has the form of the classical equation, apart from
the extra term
\begin{equation}
Q(\overrightarrow{x},t)=-\frac{\hbar^{2}}{2m}\frac{\nabla^{2}R(\overrightarrow{x},t)}{R(\overrightarrow{x},t)}
\end{equation}
where function Q has been called the quantum potential
{\cite{Holland}}. But, it is clear that the calculation and
analysis of Q, by using our total wavefunction (4), is not very
simple.  On the other hand, we can use the form of Newton's second
law, in which the particle is subject to a quantum force
$(-\overrightarrow{\nabla} Q)$ in addition to the classical force
$(-\overrightarrow{\nabla} V)$ {\cite{Holland}}, namely
\begin{equation}
\overrightarrow{F}=-\overrightarrow{\nabla} (Q+V).
\end{equation}
Now, if we utilize the equation of motion of the center of mass
$y$-coordinate (18) and the equation (20), we shall obtain the
quantum potential for the center of mass motion $(Q_{cm})$. Thus,
we can write
\begin{equation}
-\frac{\partial Q}{\partial x}=m\ddot{x}=0
\end{equation}
\begin{equation}
-\frac{\partial Q}{\partial
y}=m\ddot{y}=\frac{my_{0}(\hbar/2m\sigma_{0}^{2})^{2}}{(1+(\hbar
t/2m\sigma_{0}^{2})^{2})^{3/2}}=\frac{my_{0}^{4}}{y^{3}}(\frac{\hbar}{2m\sigma_{0}^{2}})^{2}
\end{equation}
where the result of equation (21) is clearly due to motion of
plane wave in the $x$-direction.  In addition, we assume that
$\nabla V=0$ in our experiment.  Thus, our effective quantum
potential is only a function of $y$-variable and it has the form
\begin{equation}
Q=\frac{my_{0}^{4}}{2y^{2}}(\frac{\hbar}{2m\sigma_{0}^{2}})^{2}=\frac{1}{2}my_{0}^{2}\frac{\hbar/2m\sigma_{0}^{2})^{2}}{1+(\hbar
t/2m\sigma_{0}^{2})^{2}}.
\end{equation}
If $y_{0}=0$, the quantum potential for the center of mass of the
two particles is zero at all times and it remains on the $x$-axis.
However, if $y_{0}\neq 0$, then the center of mass cannot touch or
cross the $x$-axis.  These conclusions are consistent with our
earlier result (eq. (18)).

\section{SQM forecast and its comparison with BQM}

So far, we have been studying the results obtained from BQM at the
individual level.  But it is well known from SQM that the
probability of simultaneous detection of two particles at $y_{M}$
and $y_{N}$, at the screen $S_{1}$ and $S_{2}$ is equal to
\begin{equation}
P_{12}(y_{M},y_{N})=\int_{y_{M}}^{y_{M}+\triangle}dy_{1}\int_{y_{N}}^{y_{N}+\triangle}dy_{2}|\psi(x_{1},y_{1};x_{2},y_{2};t)|^{2}.
\end{equation}
The parameter $\Delta$, which is taken to be small, is a measure
of the size of the detectors.  It is clear that the probabilistic
prediction of SQM is in disagreement with the symmetrical
prediction of BQM, because SQM predicts that probability of
asymmetrical detection at the individual level of pair of
particles can be different from zero, in opposition to BQM's
symmetrical predictions.  In addition, based on SQM's prediction,
the probability of finding two particles at one side of the
$x$-axis can be nonzero while we showed that BQM's prediction
forbids such events in our experiment.  In other words, its
probability must be exactly zero.  Thus, if we provide necessary
arrangements to perform this experiment, we must abandon one of
the two theories or even both as a complete description of the
universe.

Now the question arises as to whether this difference persists if
we deal with an ensemble of pair of particles? To answer this
question, we consider an ensemble of pair of particles that have
arrived at the detection screens $S_{1}$ and $S_{2}$ at different
times $t_{i}$.  It is well known that, in order to ensure the
compatibility between SQM and BQM for ensemble of particles, Bohm
added a further postulate to his three basic and consistent
postulates {\cite{Bohm,Holland}}.  Based on this further
postulate, the probability that a particle in the ensemble lies
between $\overrightarrow{x}$ and
$\overrightarrow{x}+d\overrightarrow{x}$ at time $t$ is given by
\begin{equation}
P(\overrightarrow{x},t)=R^{2}(\overrightarrow{x},t).
\end{equation}
Thus, using BQM, the probability of simultaneous detection for all
pairs of particles of the ensemble arriving at the two screens at
different instant of time $t_{i}$, with $y_{0}=0$, is
\begin{eqnarray}
P_{12}=\lim_{N\rightarrow\infty}\sum_{i=1}^{N}R^{2}(y_{1}(t_{i}),-y_{1}(t_{i}),t_{i})\equiv\int^{+\infty}_{-\infty}dy_{1}\int^{+\infty}_{-\infty}dy_{2}|\psi(y_{1},y_{2},t)|^{2}=1
\end{eqnarray}
where every term in the sum shows only one pair arriving on the
screens $S_{1}$ and $S_{2}$ at the point
$(y_{1}(t_{i}),-y_{1}(t_{i}))$ and time $t_{i}$, weighted by the
corresponding density $R^{2}$.  If all times of $t_{i}$ are taken
to be $t$, then the summation on $i$ can be changed to an integral
over all paths that cross the screens $S_{1}$ and $S_{2}$ at that
time.  Now, we can consider the joint probability of two points
$y_{M}$ and $y_{N}$ on the two screens at time $t$ that are not
symmetric about the $x$-axis, but we know that they are not
detected simultaneously.  Then, one can obtain the probability of
detecting two particles at two arbitrary points $y_{M}$ and
$y_{N}$
\begin{equation}
P_{12}(y_{M},y_{N})=\int_{y_{M}}^{y_{M}+\Delta}dy_{1}\int_{y_{N}}^{y_{N}+\Delta}dy_{2}|\psi(y_{1},y_{2},t)|^{2}
\end{equation}
which is similar to the prediction of SQM (eq. (24)) but obtained
in a Bohmian way {\cite{Ghose}}.  Thus, it appears that for a
geometrically symmetric arrangement, the possibility of
distinguishing the two theories at the statistical level is
impossible, as was expected [1-3, 9, 17].

\section{Selective detection and comparison of SQM with BQM at the statistical level}

In the previous section, we have shown that SQM and BQM have
different predictions for our suggested experiment, at the
individual level.  Since SQM talks about individual events in
probabilistic terms, the existence of different predictions by the
two theories at the individual level is not a strange result.  On
the other hand, we have seen that the two theories, for a
geometrically symmetric arrangement, are consistent at the
ensemble level.  Here, one can ask whether the individual level is
the only area to distinguish between the two theories and whether
the disagreement between them cannot
 appear at the ensemble level. In this section, we answer this question in negative, and
 we shall provide
conditions under which SQM can be interpreted as a vague theory at
the ensemble level.

\subsection{The case where $y_{0}$ is exactly zero}

We have seen that the assumption of $y_{0}=0$ is not in
contradiction with the statistically results of SQM and, in
consequence, with experiment.  Thus, we can assume that initially
each particle in the source is statistically distributed according
to the absolute square of the wavefunction, but, this distribution
is completely symmetric so that the $y$-coordinate of the center
of mass is on the $x$-axis.  If we can prepare such a special
source with two correlated particles, we can try to do our
experiment in the following fashion: particles emitted from the
source $S$ into the right hand side of the experimental set-up can
pass through slits $A$ or $B$.  Since we have assumed that the
total momentum of the pair of particles is zero, if one of the
particles goes through the slit $A$ for instance, the other
particle must go through the slit $B^{'}$ on the left hand side of
the experimental set-up.  Based on BQM and using equation (16), we
infer that the particle passing through the slit $A$ must be
detected on the upper half plane of the $x$-axis on the screen
$S_{1}$.  The same thing must occur for the other particle that
passes through $B^{'}$, but in the lower half plane of the
$x$-axis on the $S_{2}$ screen. Using this prediction, we assume
that only those particles arriving at $S_{2}$ are recorded for
which there is a simultaneous detection of the other particle at
the upper side on $S_{1}$.  We called this special detection, in
which some of the selected particles are recorded, a selective
detection.  Thus, based on the prediction power of BQM, we will
record two particles symmetric with respect to the $x$-axis, for
each emitted pairs of particles.  If we wait to record an ensemble
of particles, we will see an interference pattern of particles on
the lower half plane of the $S_{2}$ screen.  On the other hand,
based on SQM, the probability of finding a particle at any point
on the $S_{2}$ screen, even at the upper side, is nonzero and
there is no compulsion to detect pairs of particles symmetrically
on the two sides of the $x$-axis, as it can be seen from equation
(24) and is depicted in Fig. 1.  Therefore, if we accept that SQM
is still efficient and unambiguous for the selective detection,
the interference pattern will be seen on the whole screen $S_{2}$,
particularly at the upper side of it, at the ensemble level.
Consequently, we shall have observable results to distinguish the
two theories, SQM and BQM.

\subsection{The case where $y_{0}$ is statistically distributed}
One can argue that $y_{0}$ cannot have a well-defined position and
it must be distributed according to Born's principle. However, we
will show that this objection cannot alter our obtained results.
Assume that, $\langle y_{0}\rangle=0$ but $\triangle y_{0}\neq 0$.
If we provide conditions in which $\triangle y_{0}$ is very small
and $\hbar t/2m\sigma_{0}^{2}\sim 1$, we can still detect
particles symmetrically with respect to the $x$-axis, with a good
approximation.  To obtain symmetrical detection about the $x$-axis
with reasonable approximation, the center of mass variation from
the $x$-axis must be smaller than the distance between any two
neighboring maxima, that is
\begin{equation}
y\ll\frac{\lambda D}{2Y}\simeq\frac{\pi\hbar t}{Ym}
\end{equation}
where $\lambda$ is the de Broglie wavelength.  For conditions
$\hbar t/2m\sigma_{0}^{2}\sim 1$, $Y\sim\sigma_{0}$ and using
equation (18), one can obtain
\begin{equation}
y_{0}\ll\frac{\pi\hbar t}{Ym}\sim\sigma_{0}.
\end{equation}
Therefore, if we use a source with $\triangle y_{0}\ll\sigma_{0}$,
we shall obtain $y\simeq y_{0}\ll\sigma_{0}$ for each individual
observation, and our symmetrical detection can be maintained with
a good approximation.  It is evident that, if one considers
$\triangle y_{0}\sim\sigma_{0}$, as was done in
{\cite{Marchildon}}, the incompatibility between the two theories
will be disappeared. But, we believe that, instead of the usual
one-particle two-slit experiment with $\triangle
y_{0}\sim\sigma_{0}$, our correlated two-particle system provides
a new situation in which we can adjust $y_{0}$ independent of
$\sigma_{0}$, so that
\begin{equation}
y_{0}=\frac{1}{2}(y_{1}+y_{2})_{t=0}\ll\sigma_{0}.
\end{equation}
Although it is obvious that $(\triangle y_{1})_{t=0}=(\triangle
y_{2})_{t=0}\sim\sigma_{0}$, but position correlation between the
two entangled particles cause that they always satisfy equation
(30).  Furthermore, if it is assumed that $y_{0}$ is statistically
distributed, another problem can be raised, which is mentioned by
Marchildon {\cite{Marchildon}}.  We have shown that, if both
particles are simultaneously on the $x$-axis, both velocities in
the $y$-direction vanish, and neither particle could cross or be
tangent to the $x$-axis.  However, under $\triangle y_{0}\neq 0$
condition, pairs of particles cannot be simultaneously on the
$x$-axis and we have not the aforementioned constraint on the
motion of particles (relations (16)).  However, using our
selective detection, we can still obtain our last result, because
the center of mass of the two particles are on the $x$-axis with
reasonable approximation.  It is clear that, under such a
condition, one cannot claim that the particle detected at the
upper (lower) side must have passed through upper (lower) slit, in
spite of $y_{0}=0$ condition.  To confirm these results, it is
worth noting that, Durr {\it{et al.}} {\cite{Durr}} argue that the
selective detection can alter the statistical predictions of the
two theories:``{\it{note that by selectively forgetting results we
can dramatically alter the statistics of those that we have not
forgotten.  This is a striking illustration of the way in which
Bohmian mechanics does not merely agree with the quantum
formalism, but, eliminating ambiguities, clarifies, and sharpens
it.}}".  Elsewhere {\cite{Golshanii}}, we have utilized another
kind of selective detection by which we could alter statistical
prediction of SQM, using BQM for an interference device that
contains two unentangled particles.

\section{Conclusion}

In this investigation, we have suggested an experiment to
distinguish between SQM and BQM.  In fact, we believe that some
particular experiments that for one reason or another have not yet
been performed can decide between them.  Thus, it has been shown
that a two double-slit experiment set-up, along with a source of
two identical non-relativistic particles with total momentum of
zero, emitted at suitable time intervals, has the following characteristics:\\
1) The suggested experiment will yield different observable
predictions for SQM and BQM, at the individual level.\\
2) The two theories yield the same interference pattern at the
ensemble level without using a selective detection, as is expected.\\
3) Since in BQM the particles are distinguishable and their past
history are known, using selective detection, it has been shown
that either the two theories will predict different results at the
statistical level, or that BQM has more predictive power than SQM.
It is shown that selective detection can be considered as a tool
for arriving to a new realm in which trajectory interpretation is
sharply formulated, while the standard interpretation is ambiguous
and silent even at the ensemble level.

Therefore, it seems possible to distinguish between the two
theories and to see whether BQM is a worthy successor to SQM.


\begin{thebibliography}{Dillo 83}
\bibitem{Bohm} D. Bohm, Part I, {\it{Phys. Rev.}} {\bf{85}} 166 (1952); Part II, {\bf{85}} 180 (1952).
\bibitem{Holland} P.R. Holland, {\it{The Quantum Theory of Motion}}, Cambridge
University Press, Cambridge, 1993.
\bibitem{Cushing} J.T. Cushing, {\it{Quantum Mechanics: Historical Contingency and the Copenhagen Hegemony}}, The University of
Chicago Press, Ltd., London, 1994.
\bibitem{Bell} J.S. Bell, {\it{Speakable and Unspeakable in Quantum
Mechanics}}, Cambridge University Press, Cambridge, 1987.
\bibitem{Abolhasani} M. Abolhasani and M. Golshani, {\it{Phys. Rev. A}} {\bf{62}} 12106
(2000).
\bibitem{Englert} B.G. Englert, M.O. Scully, G. Sussman and H. Walther, {\it{Z.
Naturf.}} {\bf{47}}a  1175 (1992).
\bibitem{Scully} M.O. Scully, {\it{Phys. Scripta}}, T{\bf{76}} 41 (1998).
\bibitem{Dewdney} C. Dewdney, L. Hardy and E.J. Squires, {\it{Phys. Lett. A}}
{\bf{184}} 6 (1993).
\bibitem{Hiley} B.J. Hiley, R.E. Callaghan and O.J.E. Maroney, quant-ph/0010020.
\bibitem{Vigier} J.P. Vigier, {\it{Phys. Lett. A}} {\bf{270}} 221 (2000).
\bibitem{Ghose} P. Ghose, quant--ph/0001024 and quant-ph/0003037.
\bibitem{Golshani} M. Golshani and O. Akhavan, quant-ph/0009040.
\bibitem{Golshanii} M. Golshani and O. Akhavan, quant-ph/0103100.
\bibitem{Ghosee} P. Ghose, quant-ph/0102131.
\bibitem{Marchildon} L. Marchildon, quant-ph/0101132.
\bibitem{Belousek} D.W. Belousek, {\it{Found. Phys.}} {\bf{30}} 153 (2000).
\bibitem{Durr} D. Durr, S. Goldstein and N. Zanghi, {\it{J. Stat. Phys.}}
{\bf{67}} 843 (1992).
\end{thebibliography}
\end{document}